\documentclass{llncs}
\usepackage{multirow}
\usepackage{amsmath}
\usepackage[table]{xcolor}
\usepackage{graphicx}
\usepackage{amssymb}
\usepackage{wrapfig}
\usepackage{hyperref}
\usepackage{pbox}

\def\phi{\varphi}
\def\theta{\vartheta}

\def\ff{\mathit{false}}
\def\tt{\mathit{true}}

\def\LTLSAT{{Polsat}}
\def\Aalta{{Aalta}}

\title{\LTLSAT: A \underline{Po}rtfolio \underline{L}TL \underline{Sat}isfiability Solver}
\author{Jianwen Li\inst{1} \and
Geguang Pu\inst{1} \and
Lijun Zhang\inst{2} \and \\
Yinbo Yao\inst{1} \and
Moshe Y. Vardi\inst{3} \and
Jifeng He\inst{1}
}

\institute
{
  \inst{}%
Shanghai Key Laboratory of Trustworthy Computing, East China Normal University, 
 P. R.China
  \and
  \inst{}%
State Key Laboratory of Computer Science, Institute of Software, Chinese Academy of Sciences
\and
\inst{}
Department of Computer Science, Rice University, USA
}

\begin{document}

\maketitle

\begin{abstract}
  In this paper we present a portfolio LTL-satisfiability solver,
  called \LTLSAT. To achieve fast satisfiability checking for LTL
  formulas, the tool integrates four representative LTL solvers: pltl,
  TRP++, NuSMV, and \Aalta. The idea of \LTLSAT\ is to run the
  component solvers in parallel to get best overall performance; once
  one of the solvers terminates, it stops all other solvers.
  Remarkably, the \LTLSAT\ solver utilizes the power of modern
  multi-core compute clusters.  The empirical experiments show that
  \LTLSAT\ takes advantages of it. Further, \LTLSAT\ is also a testing
  platform for all LTL solvers.
\end{abstract}

\section{Introduction}
Linear Temporal Logic (LTL) \emph{satisfiability checking} plays an
important role in ensuring the quality of temporal specifications that
are often used in an early stage in designing processes \cite{RV10}.
Temporal system requirements consist of a set of LTL properties
identifying system properties that are  supposed to hold in all system
executions.  Thus, these formulas must be satisfiable, and their conjunction
must be satisfiable as well.  Satisfiability checking must be scalable
due to the the need to handle complex temporal properties.

Earlier work \cite{RV10} and \cite{SD11} reported on
extensive experimental investigations in LTL satisfiability checking.
Rozier and Vardi reached the conclusion that when it comes to
LTL satisfiability checking via reduction to model checking, the symbolic
approach is superior to the explicit approach \cite{RV10}.
Nevertheless, they showed in later work that no single symbolic approach
is dominant across their extensive benchmark suite \cite{RV11}.
Schuppan and Darmawan considered a wide range of solvers
implementing three major classes of algorithms: reduction to model
checking, tableau-based approaches, and temporal resolution \cite{SD11}.
They argued that no solver dominates across their benchmark suite. Our previous
work~\cite{LZPVH13} on LTL satisfiability checking supports this conclusion further,
but discovers that on-the-fly explicit approach is advantageous in
checking satisfiable formulas. This motivated us to extend the portfolio
approach of \cite{RV11}, but go beyond symbolic model-checking techniques and
develop a portfolio LTL satisfiability solver that integrates
several types of LTL satisfiability solvers and utilizes the power of modern
multi-core compute clusters.

We describe here a portfolio LTL satisfiability solver, called
\LTLSAT%
\footnote{The tool can be download at
  \url{http://www.lab205.org/ltlsat}}.  The tool integrates four
representative LTL solvers: pltl, TRP++, NuSMV, and \Aalta. The
approach of \LTLSAT\ is to run the solvers in parallel to get the best
overall performance; once one of the solvers terminates, it stop all
other solvers. To test the performance of \LTLSAT, we collect in this
paper all existing benchmarks of LTL satisfiability checking
\cite{RV10} \cite{SD11} \cite{LZPVH13}.  

The empirical results show that \LTLSAT\ takes advantages of the
integrated solvers, and scales better for a large selection of
benchmarks, especially those random formulas.

Another contribution of this paper is that \LTLSAT\ provides testing
platform for LTL solvers. A tool developer can use the benchmarks
provided by the platform to test the solver under development and
compare the results with other solvers. Thus, the tool developer can
study carefully the advantage and disadvantage of the tool under
development, and optimize it based on the testing results. For
instance, our earlier tool, \Aalta, benefited from this platform by
designing new heuristics to improve tool performance.

\section{Solvers}
\cite{SD11} classified three major classes of solvers based on the
techniques the solvers often use: \emph{reduction to model checking},
\emph{tableau-based approaches} and \emph{temporal resolution}. Here,
we add a new class named \textit{hybrid approaches}, which combines
different techniques together to achieve better performance. Solvers
selection strategy is discussed below.

\noindent\textit{\bf Reduction to model checking.} We choose NuSMV
\cite{CCGGPRST02} as the representative. \cite{RV10} and ~\cite{SD11}
carefully evaluated model checking tools such as NuSMV and ALASKA
\cite{DDMR08}.  Based on their observation, we ruled out explicit
state model checkers, as they did not scale comparing to symbolic
ones. ALASK is not included because it fails to run on our
experimental cluster platform. Thus NuSMV is chosen with both its BDD-
and SAT-based appraoches.

\begin{figure}
\begin{minipage}[b]{0.45\linewidth}
\centering
\includegraphics[scale=0.35]{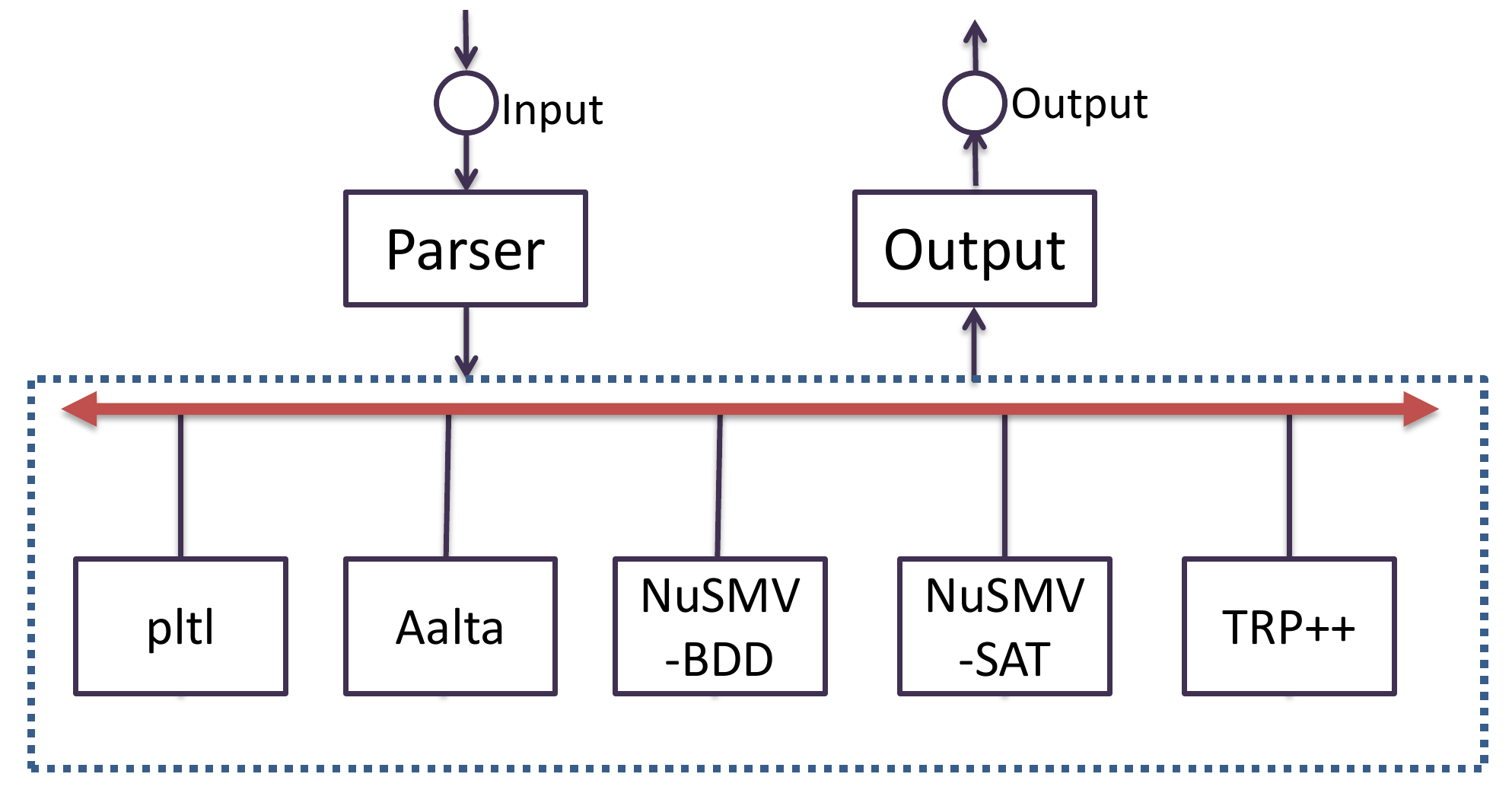}
\caption{The Framework of \LTLSAT}\label{fig:framework}
\end{minipage}
\hspace{1cm}
\begin{minipage}[b]{0.45\linewidth}
\centering
\scalebox{0.7}{
  \begin{tabular}{|c|c|}
    \hline
      operators  &  symbols \\
      \hline
      $\neg$  &  $!$, $\sim$ \\
      \hline
      $\wedge$  &  \& , \&\& \\
      \hline
      $\vee$  &  $|$, $||$ \\
      \hline
      $X$  &  X \\
      \hline
      $U$  &  U \\
      \hline
      $R$  &  R , V \\
      \hline
      $G$  &  G, [] \\
      \hline
      $F$  &  F, $\langle\rangle$ \\
      \hline
      $\rightarrow$  &  $\rightarrow$\\
      \hline
      $\leftrightarrow$  &  $\leftrightarrow$ \\
      \hline
      $\tt$  &  true, TRUE \\
      \hline
      $\ff$  &  false, FALSE \\
    \hline
  \end{tabular}
}
\caption{Logic Operators in \LTLSAT}\label{tab:operator}
\end{minipage}
\end{figure}

\noindent\textit{\bf Tableau-based approaches.} We choose pltl~\cite{Sch98}
as the representative. From the experiments by \cite{SD11}, pltl has
the best potential in this type of solvers. Our previous experiments
also confirm this conclusion.

\noindent\textit{\bf Temporal resolution.} We choose TRP++~\cite{HK03} as the representative. From the observations of \cite{SD11}, TRP++ dominates most of cases in this type of solvers.

\noindent\textit{\bf Hybrid approaches.} We choose \Aalta~\cite{LZPVH13} as the representative. This type of solvers includes PANDA~\cite{RV11} and \Aalta~\cite{LZPVH13}. PANDA tool is basically a model checking based approach but integrates multiple novel encodings of symbolic transition-based B\"uchi automata. \Aalta\ belongs to the tableau-based approach but integrates some interesting heuristics. Our previous study showed that \Aalta\ has a best potential in most cases compared to PANDA.

Summarizing, \LTLSAT\ tool integrates solvers including NuSMV, pltl,
TRP++ and \Aalta. Since NuSMV provides both the BDD-based and
SAT-based model checking, we integrated both two functionalities in
\LTLSAT\ respectively.

\section{The framework of \LTLSAT}

A general framework of \LTLSAT\ is shown in Fig. \ref{fig:framework}:
it consists of three components, that are, the input, solver set and
output module. Details for each component are specified in the
following.

As soon as \LTLSAT\ is invoked, it creates five threads to run these
solvers -- each solver occupies one unique thread. Once one of the
solvers finishes checking then the corresponding thread will kill all
other threads, which is illustrated in the figure, as \emph{all solvers
  can communicate through the bus}. After that, the remaining thread
will send the solver's results to the Output module for further
processing.

One of main \LTLSAT's features is, it also supports to integrate
external solvers in addition to those have been integrated -- with the
only restriction that the solver has to provide the same input and
output interface as \LTLSAT.  Using the parameter
\verb+-add solverpath+, one can import an external solver whose path
is located in \verb+solverpath+.  This feature makes \LTLSAT\
extensible, and provides testing platform for LTL solvers.

\begin{table}[h]
\centering
\scalebox{0.8}{
\begin{tabular}{|l|l|l|l|l|l|l|}
  \hline
  Formula Type  &  pltl  &  TRP++  &  \pbox{2cm}{NuSMV\\-BDD}  &  \pbox{2cm}{NuSMV\\-BMC}  &  \Aalta  & \LTLSAT\\
\hline
/acacia/demo-v3  & 366.805  &  5.958  &  2753.55  &  1.004  &  557.862  &  4.326\\
\hline
/alaska/lift  &  5800.595  &  14989.337  &  13478.447  &  2797.13  &  8151.248  &  \cellcolor{blue!25}2721.996\\
\hline
/anzu/amba  &  965.456  &  5914.278  &  6088.505  &  398.177  &  2278.774  &  410.652\\
\hline
/anzu/genbuf  &  2849.786  &  6609.282  &  7085.315  &  695.145  &  2405.892  &  697.07\\
\hline
/Rozier/counter  &  1415.379  &  1570.318  &  5639.615  &  3981.308  &  3771.958  &  \cellcolor{blue!25}1388.318\\
\hline
/Rozier/formulas  &  364.475  &  50066.122  &  3918.415  &  6663.472  &  463.271  &  \cellcolor{blue!25}232.728\\
\hline
/Rozier/pattern  &  15.13  &  5530.001  &  17644.459  &  31.484  &  28.592  &  34.332\\
\hline
/schuppan/O1formula  &  1026.916  &  1148.885  &  2058.842  &  1626.036  &  6.114  &  5.739\\
\hline
/schuppan/O2formula  &  1082.35  &  1591.756  &  2167.806  &  1622.142  &  6.447  &  7.359\\
\hline
/schuppan/phltl  &  900.997  &  1810.009  &  1355.264  &  1081.194  &  1102.993  &  \cellcolor{blue!25}725.139\\
\hline
/trp/N5x  &  14.44  &  12575.152  &  12.681  &  6546.099  &  1356.775  &  31.627\\
\hline
/trp/N5y  &  2761.521  &  8933.292  &  1395.545  &  2763.737  &  2766.555  &  \cellcolor{blue!25}1374.437\\
\hline
/trp/N12x  &  20572.63  &  34345.41  &  25878.431  &  10513.257  &  2319.798  &  2387.982\\
\hline
/trp/N12y  &  4127.099  &  22231.442  &  22807.655  &  4026.722  &  4033.153  &  4042.285\\
\hline
Total  &  44506.513  &  169667.268  &  112307.805  &  44250.091  &  30459.832  &  15332.828\\
\hline

\end{tabular}
}
\caption{Comparison results for the \textit{Schuppan-collected} benchmarks}\label{tab:schuppan}
\end{table}

\subsection{Input}

\LTLSAT\ supports the standard LTL syntax, that is, an LTL formula $\phi$ is defined recursively
as:

$\phi\ ::=\ \tt\ |\ \ff\ |\ p\ |\ \neg\ \phi\ |\ \phi\ \wedge\ \phi\ |\ \phi\ \vee\ \phi\ |\ X\phi\ |\ \phi\ U\ \phi$;

Also, we can introduce the operator $R$ (release), which is the dual
operator of $U$ (until): $\phi_1 R \phi_2\equiv \neg (\neg\phi_1 U
\neg\phi_2)$. Specially, the $G$ (Global) and $F$ (Future) operators
are interpreted as $G\phi\equiv \ff R\phi$ and $F\phi\equiv \tt
U\phi$. As the same in propositional logic, it still holds that
$\phi_1\rightarrow\phi_2\equiv\neg\phi_1\vee\phi_2$ and
$\phi_1\leftrightarrow\phi_2\equiv(\neg\phi_1\vee\phi_2)\wedge
(\phi_1\vee\neg\phi_2)$ for LTL formulas.  Among the operators above,
\LTLSAT\ recognizes the alternative symbols. The explicit representing
is shown in Table~\ref{tab:operator}.

\LTLSAT\ has integrated several off-the-shelf solvers, and these
solvers may have different input formats. To successfully invoke these
solvers, the Parser module also integrates internal translators from
the input of \LTLSAT\ to those of them.

\subsection{Output}
The output of \LTLSAT\ includes the following information: the checking
result (``sat'' or ``unsat''), the solver where the result comes from, and
the execution eclipse time. As the outputs vary on
the different solvers, the Output module shown in the
Fig. \ref{fig:framework} is designed to unify the outputs from
different integrated solvers.

\section{Empirical Experiments}

We conducted all the experiments on SUG@R cluster\footnote{\url{http://www.rcsg.rice.edu/sharecore/sugar/}}. SUG@R is comprised of 134 Sun Microsystems SunFire x4150 nodes, each of which contains two quad-core 2.83GHz Intel Xeon Harpertown CPUs with 16GB RAM.

\begin{figure}
\begin{minipage}[b]{0.45\linewidth}
\centering
\includegraphics[scale = 0.53]{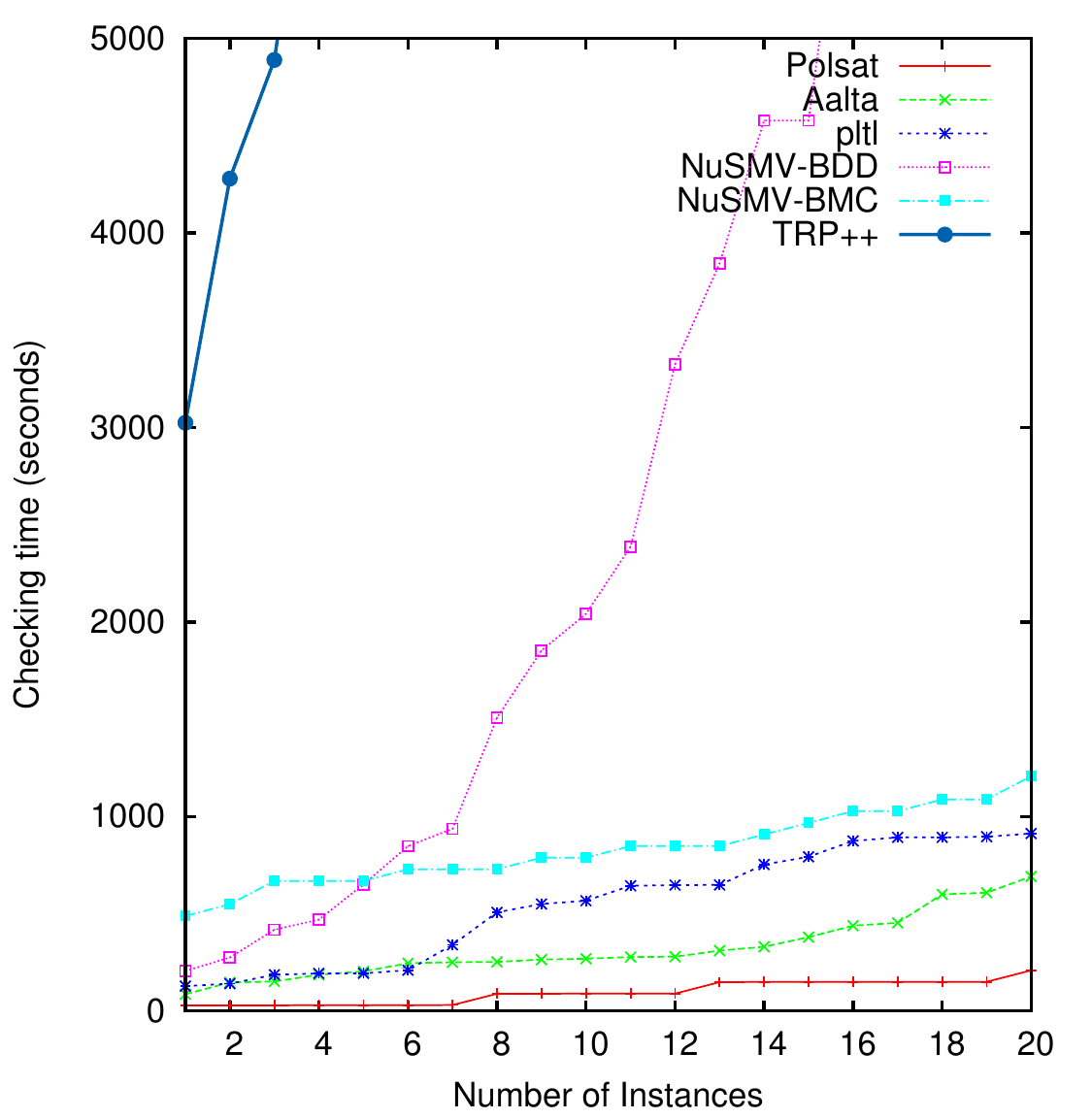}
\caption{Experimental results on extended random formulas with 3 variables.}
\label{fig:random}
\end{minipage}
\hspace{0.6cm}
\begin{minipage}[b]{0.45\linewidth}
\centering
\includegraphics[scale = 0.53]{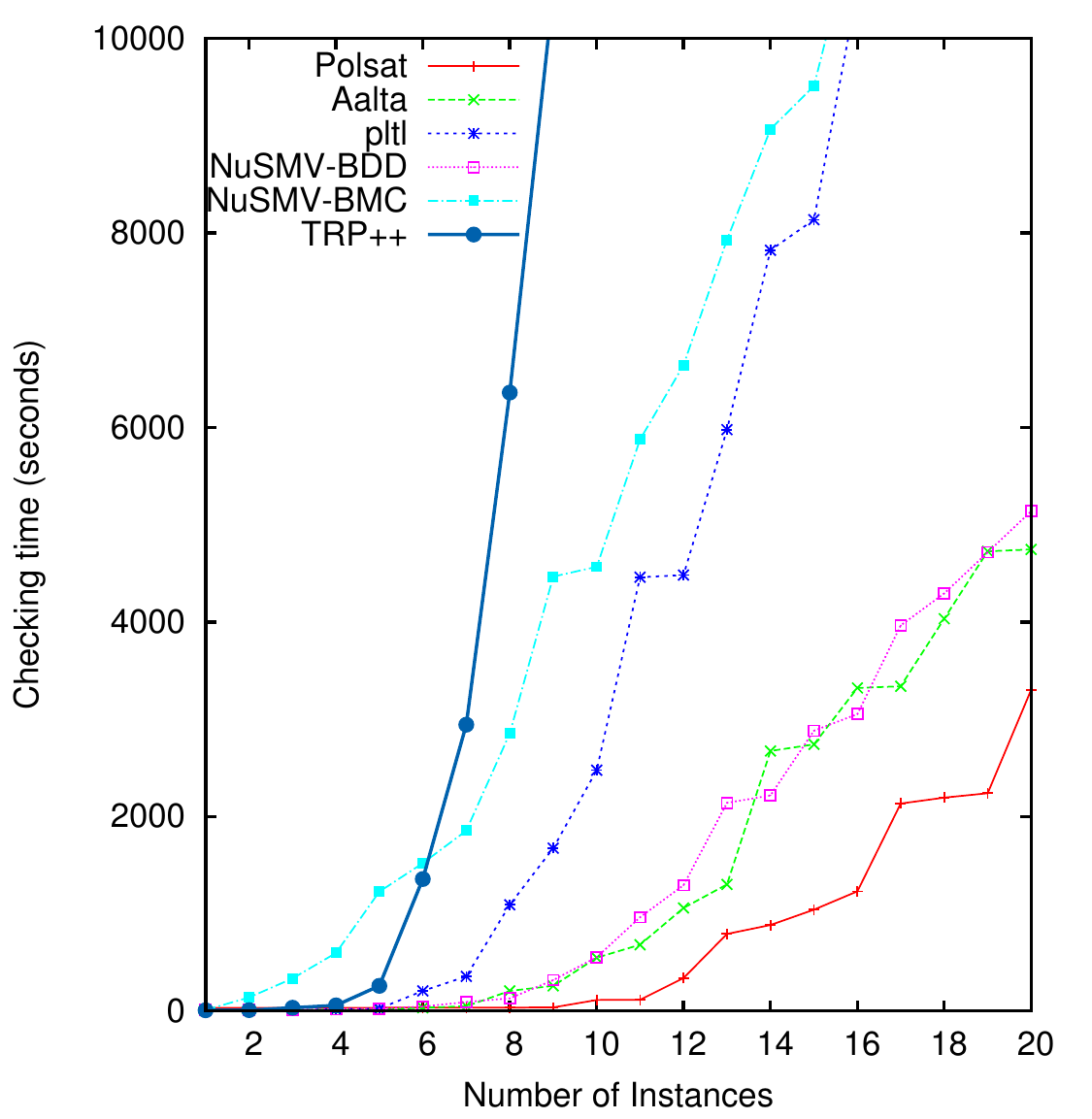}
\caption{Experimental results on random conjunction formulas.}
\label{fig:random_conjunction}
\vspace{0.3cm}
\end{minipage}
\end{figure}

The benchmarks we used are mainly from \cite{SD11}. We call the benchmarks
\textit{Schuppan-collected}
for convenience.
To check the scalability of LTL solvers, we
also tested the \textit{random conjunction formulas} proposed
in~\cite{LZPVH13}.  A random conjunction formula has the form of
$\bigwedge_{1\leq i\leq n} P_i$, where $P_i$ is a random specification
pattern\footnote{\url{http://patterns.projects.cis.ksu.edu/documentation/patterns/ltl.shtml}}.
In our experiments, the timeout
for every testing formula is 60 seconds. Note the time is also counted
if the running time of a formula checking reaches the timeout.

The experimental results on \textit{Schuppan-collected} benchmarks are
shown in Table \ref{tab:schuppan}. The first row lists all the types
of this benchmark and the second to seventh ones list the total
execution time for the corresponding type of formulas.  Theoretically
speaking, \LTLSAT\ should be always the best. But as seen from the
table, there may be some deviations between the results from \LTLSAT\
and those best from integrated tools. This is due to the overhead we
have to pay on pre-processing the input formula for each integrated
tool (different tools have different input formats). In the table we
also highlight the benchmarks for which \LTLSAT\ is faster than the
best of all solvers.  The reason is that individual solver may not be
superior to all cases in one type formulas while \LTLSAT\ gets the
best from different solvers in the same type, which leads to better
overall performance for some benchmarks.

To show the power of \LTLSAT\ on hard problems, we present the
experimental results on two type of formulas. First, we extend
Rozier's random formulas~\cite{RV10} in \textit{Schuppan-collected}
benchmark via enlarging the size of generated formulas, and choosing
500 cases for each size (the formula length from 100 to 200). Second,
we test the \textit{random conjunction formulas} varying on the number
of conjunctions (1-20) and select 500 random cases for each
conjunction. The experimental results are shown in
Fig. \ref{fig:random} and Fig. \ref{fig:random_conjunction}. In the
figures we use the cactus plot to show the relationship between the
number of instances solved by tools (x-axis) and their total checking
costs (y-axis, with the second unit). One can see clearly from the
figures that \LTLSAT\ solves more cases with the same time, and has
the best overall performance for these benchmarks.

As the integrated tool of off-the-shelf solvers, \LTLSAT\ also
provides a platform for competitions of LTL satisfiability solvers. By
observing the best result among different solvers, \LTLSAT\ knows
which solver performs best for a given type of formulas. For example,
for the /alaska/lift formulas, the NuSMV-BMC performs best; \Aalta\
does the best job for /schuppan/O1formula and /schuppan/O2formula
formulas. The other benefit of \LTLSAT\ is to make integrated tools
potentially to optimize their performances by utilizing the
experimental results.

\section{Conclusion}
We present a portfolio LTL satisfiability checker as well as an LTL
testing platform, \LTLSAT, by integrating existing off-the-shelf LTL
satisfiability solvers. The goal is to provide a best LTL
satisfibility solver by fully exploiting the distributed/multicore
systems. Our empirical experimental results show that \LTLSAT\ can
have good overall performance for many benchmarks.

\bibliographystyle{alpha}
\bibliography{ok,cav}

\newpage
\appendix

\section{A Simple Demonstration}

In this section we show how to use \LTLSAT\ by a simple demonstration. We will explain the input and output information of
\LTLSAT\ as well as the parameters the tool provides.

\LTLSAT\ is run on Linux or Unix operating systems. After successfully installed \LTLSAT\
and all its integrated solvers, one can directly type ``./polsat'' in the shell command line. By default the following information will show up:

\begin{verbatim}
  please input the formula:

\end{verbatim}

This means \LTLSAT\ is waiting for the input. After you type the formula, such as ``a U b'' in
the shell, then \LTLSAT\ will produce the following output information:

\begin{verbatim}
  sat
  from pltl
  eclipse time: 0.001s
\end{verbatim}

The first line tells that the input formula is satisfiable; The second line shows this result is
from pltl solver; The third line displays the checking time  is 0.001 seconds.

Alternatively, for the same case, one can directly type ``./polsat ``a U b'''' in the command line,
and will get the same result.

\subsection{Evidence for the Satisfiable formula}

Similar to most of existed LTL satisfiability solvers, \LTLSAT\ provides an interface to show an
``evidence'' for the satisfiable formula. By using the same formula ``a U b'', if one uses the
parameter ``-e'' of \LTLSAT, that means, type ``./polsat -e ``a U b'''' in the command line, then
the output becomes:

\begin{verbatim}
  sat
  (b)
  from Aalta
  eclipse time: 0.002s
\end{verbatim}

Here ``(b)'' in the second line represents the infinite trace $b^\omega$: obviously
$b^\omega\models a U b$ holds. When the input formula is unsatisfiable, the flag ``-e'' will be
ignored. Note here that not all integrated solvers provide the evidences for satisfiable formulas,
so \LTLSAT\ is designed to get the evidences from the \Aalta\ solver since this solver has the functionality.

\section{ Examples}

The motivation of \LTLSAT\ comes from that, none of existed LTL satisfiability solvers perform best
for all benchmarks. In other words, each solver has its own advantages on some kind of formulas. The implementation
of \LTLSAT\ confirms that it inherits all advantages of integrated solvers. In the
following we show two cases. Since small formulas do not make large derivations
among solvers, we choose the formulas of large size as the demonstration.

\subsection{NuSMV-BMC performs best on lift formulas}

The lift formulas is one benchmark from \textit{Schuppan-collected} for the lift specification.
The following lists one formula for the lift for three floors:

\begin{verbatim}
G(((f0 -> (!(f1) & !(f2))) & (f1 -> !(f2)))) &
!(u) & f0 & !(b0) & !(b1) & !(b2) & !(up) &
G((u <-> !(Xu))) &
G(((u -> ((f0 <-> X(f0)) & (f1 <-> X(f1)) & (f2 <-> X(f2)))) &
    (f0 -> X((f0 | f1))) & (f1 -> X((f0 | f1 | f2))) &
    (f2 -> X((f1 | f2))))) &
G(((!(u) -> ((b0 <-> X(b0)) & (b1 <-> X(b1)) & (b2 <-> X(b2)))) &
  ((b0 & !(f0)) -> X(b0)) & ((b1 & !(f1)) -> X(b1)) &
  ((b2 & !(f2)) -> X(b2)))) &
G((((f0 & X(f0)) -> (up <-> X(up))) &
   ((f1 & X(f1)) -> (up <-> X(up))) &
   ((f2 & X(f2)) -> (up <-> X(up))) & ((f0 & X(f1)) -> up) &
   ((f1 & X(f2)) -> up) & ((f1 & X(f0)) -> !(up)) &
   ((f2 & X(f1)) -> !(up)))) &
G((sb <-> (b0 | b1 | b2))) &
G((((f0 & !(sb)) -> (f0 U (sb V (F(f0) & !(up))))) &
   ((f1 & !(sb)) -> (f1 U (sb V (F(f0) & !(up))))) &
   ((f2 & !(sb)) -> (f2 U (sb V (F(f0) & !(up))))))) &
G(((b0 -> F(f0)) & (b1 -> F(f1)) & (b2 -> F(f2)))))
\end{verbatim}

Taking this formula as input, \LTLSAT\ gives the following output:

\begin{verbatim}
sat
from NuSMV-BMC
eclipse time: 0.005s
\end{verbatim}

Generally speaking, the SAT-based checking shows the best performance for lift formulas, since
the bounded model checking technique is suitable for solving  satisfiable formulas. The experiments also confirm that NuSMV-BMC performs almost best for satisfiable formulas.

\subsection{\Aalta\ performs best on /schuppan/O1formula formulas}

Let us take another example on unsatisfiable formulas. The benchmark ``./schuppan/O1formula'' formulas
are such representatives. The following shows a formula from this benchmark with the length of 100.

\begin{verbatim}
(((a1) | (b1)) & ((a2) | (b2)) & ((a3) | (b3)) &
((a4) | (b4)) & ((a5) | (b5)) & ((a6) | (b6)) &
((a7) | (b7)) & ((a8) | (b8)) & ((a9) | (b9)) &
((a10) | (b10)) & ((a11) | (b11)) & ((a12) | (b12)) &
((a13) | (b13)) & ((a14) | (b14)) & ((a15) | (b15)) &
((a16) | (b16)) & ((a17) | (b17)) & ((a18) | (b18)) &
((a19) | (b19)) & ((a20) | (b20)) & ((a21) | (b21)) &
((a22) | (b22)) & ((a23) | (b23)) & ((a24) | (b24)) &
((a25) | (b25)) & ((a26) | (b26)) & ((a27) | (b27)) &
((a28) | (b28)) & ((a29) | (b29)) & ((a30) | (b30)) &
((a31) | (b31)) & ((a32) | (b32)) & ((a33) | (b33)) &
((a34) | (b34)) & ((a35) | (b35)) & ((a36) | (b36)) &
((a37) | (b37)) & ((a38) | (b38)) & ((a39) | (b39)) &
((a40) | (b40)) & ((a41) | (b41)) & ((a42) | (b42)) &
((a43) | (b43)) & ((a44) | (b44)) & ((a45) | (b45)) &
((a46) | (b46)) & ((a47) | (b47)) & ((a48) | (b48)) &
((a49) | (b49)) & ((a50) | (b50)) & ((a51) | (b51)) &
((a52) | (b52)) & ((a53) | (b53)) & ((a54) | (b54)) &
((a55) | (b55)) & ((a56) | (b56)) & ((a57) | (b57)) &
((a58) | (b58)) & ((a59) | (b59)) & ((a60) | (b60)) &
((a61) | (b61)) & ((a62) | (b62)) & ((a63) | (b63)) &
((a64) | (b64)) & ((a65) | (b65)) & ((a66) | (b66)) &
((a67) | (b67)) & ((a68) | (b68)) & ((a69) | (b69)) &
((a70) | (b70)) & ((a71) | (b71)) & ((a72) | (b72)) &
((a73) | (b73)) & ((a74) | (b74)) & ((a75) | (b75)) &
((a76) | (b76)) & ((a77) | (b77)) & ((a78) | (b78)) &
((a79) | (b79)) & ((a80) | (b80)) & ((a81) | (b81)) &
((a82) | (b82)) & ((a83) | (b83)) & ((a84) | (b84)) &
((a85) | (b85)) & ((a86) | (b86)) & ((a87) | (b87)) &
((a88) | (b88)) & ((a89) | (b89)) & ((a90) | (b90)) &
((a91) | (b91)) & ((a92) | (b92)) & ((a93) | (b93)) &
((a94) | (b94)) & ((a95) | (b95)) & ((a96) | (b96)) &
((a97) | (b97)) & ((a98) | (b98)) & ((a99) | (b99)) &
((a100) | (b100)) & ((G c) & (X ! c)))
\end{verbatim}

The output of \LTLSAT\ is below:

\begin{verbatim}
  unsat
  from Aalta
  eclipse time: 0.04s
\end{verbatim}

This formula is unsatisfiable, and one can see it is the last term of
the formula, ((G c) \& (X ! c)), that makes the formula unsatisfiable.
If a solver provides some heuristic strategies for unsatisfiable
formulas, it can give the answer very quickly. Since \Aalta\
integrates some novel strategies to boost the search efficiency, it
performs best in this case.


\section{The Testing Integration Platform}

\LTLSAT\ is not only a portfolio LTL satisfiability solver, but also considered as a testing integration
platform for the existed or new LTL satisfiability solvers. That is to say, given the input formula, \LTLSAT\
allows all integrated solvers to run separately, and every solver will not be terminated until it
finishes checking. The \LTLSAT\ then outputs all results and eclipse time for the solvers. For
example, by adding the parameter ``-s'' and taking the following formula as the input,

\begin{verbatim}
 a & G((a -> (X(!(a)) & X(X(a))))) & !(b) & X(!(b)) &
 G(((a & !(b)) -> (X(X(b)) &
   X(((!(a) & (b -> X(X(b))) & (!(b) -> X(X(!(b))))) U a))))) &
 G(((a & b) -> (X(X(!(b))) &
   X(((b & !(a) & X(X(!(b)))) U (a | (!(a) & !(b) & X(X(b)) &
   X(((!(a) & (b -> X(X(b))) & (!(b) -> X(X(!(b))))) U a)))))))))
\end{verbatim}

which is a counter formula from the benchmark /rozier in \textit{schuppan-collected}, \LTLSAT\
gives the output:

\begin{verbatim}
  pltl: sat	0.001s
  NuSMV-BMC: sat	0.0026s
  NuSMV-BDD: sat	0.014s
  TRP++: sat	0.034s
  Aalta: sat	0.57s
\end{verbatim}

In each line of the output, it shows respectively the checking result (sat or unsat)
and the eclipse time for all solvers. With the above information, one can check whether the
checking results are consistent from all solvers, and the executing gap among different solvers. Moveover,
based on the concrete results, the tool developer may try to explore the reason of inefficiency of the tool for some benchmark, and thus optimize the tool further.

\subsection{Formulas in A File}

As a testing platform, another key functionality that \LTLSAT\ supports is to allow to input a set
of formulas stored in a file and to provide the statistics by running the integrated solvers separately. By
adding the parameter ``-sm file'' to \LTLSAT, it will read all formulas in the specified file as the inputs
 and run them separately. The final output of \LTLSAT\ in this situation
will be stored into an output file including the checking result and time for each formula. For example, when taken a set of
100 random formulas as inputs, \LTLSAT\ gives the following output:

\begin{verbatim}
  pltl		0.81s
  NuSMV-BMC	0.96s
  Aalta		1.13s
  NuSMV-BDD	2.58s
  TRP++		9.56s
  The generated file is output.txt.
 \end{verbatim}

\subsection{Adding External Solvers}

\LTLSAT\ is designed to be an open platform such that it allows to import external LTL satisfiability
solvers as well. It can be achieved by using the ``-add solverpath'' parameter of \LTLSAT. For
example, the solver ALASKA is not integrated in \LTLSAT\ currently, so we can use the flag to import it: type
``./polsat -add ``../alaska/alaska'''' in the shell command line, then the following information will
show up:

\begin{verbatim}
  ../alaska/alaska is added.
  please input the formula:
\end{verbatim}

Then the solver ALASKA is successfully added to \LTLSAT.

\end{document}